\documentclass[a4paper,11pt]{article}
\usepackage{pos}
\usepackage{orcidlink}
\usepackage{diagbox}
\usepackage{makecell, multirow}
\usepackage{pgf}
\usepackage{soul}
\usepackage{graphicx, color}
\usepackage{rotating}
\usepackage{subcaption}
\allowdisplaybreaks
\newcommand{\ket}[1]{|#1\rangle}

\definecolor{link_blue}{RGB}{51,102,204}
\hypersetup{%
pdftitle = {title},
pdfsubject = {},
pdfkeywords = {},
colorlinks = {true},
filecolor = {black},
linkcolor = {link_blue},
menucolor = {black},
citecolor = {link_blue},
urlcolor = {link_blue},
}{}

\title{Dynamics of entanglement entropy for a locally monitored lattice gauge theory}
\ShortTitle{Effects of measurements on entanglement dynamics for a lattice gauge theory}

\author*[a,b,c]{Nisa Ara}
\author[d]{Arpan Bhattacharyya}
\author[d]{Nilachal Chakrabarti}
\author[d]{Neha Nirbhan}
\author[b,c]{Indrakshi Raychowdhury}

\affiliation[a]{Department of Theoretical Physics, Tata Institute of Fundamental Research,\\Homi Bhabha Road, Mumbai 400005, India}
\affiliation[b]{Department of Physics, Birla Institute of Technology and Science Pilani, Zuarinagar, Goa 403726, India.
}
\affiliation[c]{Center for Research in Quantum Information and Technology,
Birla Institute of Technology and Science Pilani, Zuarinagar, Goa 403726, India.}

\affiliation[d]{Department of Physics, Indian Institute of Technology Gandhinagar, Gujarat 382055, India.}

\emailAdd{nisa.ara@theory.tifr.res.in}
\abstract{The $1+1$ dimensional $Z_2$ gauge theory is the simplest model that allows for quantum computation or quantum simulation to probe the fundamental aspects of a gauge theory coupled with dynamical fermions. To reliably benchmark such a system, it is crucial to understand the non-unitary quantum dynamics arising from the underlying non-Hermitian evolution and to model the effects of quantum measurements. In this work, we study post-selected filtering dynamics of physical observables for a $\mathbb {Z} _2$ gauge theory. Tensor network calculations are performed to dynamically probe entanglement entropy at larger lattice sizes. 
\textcolor{black}{We report that projective measurement of local and diagonal observables (electric and mass energy densities) in the computational basis demonstrates the absence of any measurement-induced phase transition like phenomenon, as indicated by the system-size independence of the late-time saturation value of the bipartite entanglement entropy}.}

\FullConference{The 42nd International Symposium on Lattice Field Theory (LATTICE2025)\\
2-8 November 2025\\
Tata Institute of Fundamental Research, Mumbai, India\\}
\begin{document}
\maketitle
% ------------------------------------------------------------------

\section{Introduction}\label{section-1}

The study of entanglement in lattice gauge theories (LGTs) sits at the confluence of quantum 
information, condensed matter, and high-energy physics. The $1+1$D $\mathbb{Z}_2$ gauge theory, 
the simplest foundational LGT, serves as an ideal theoretical laboratory for phenomena such as 
confinement and string breaking~\cite{RevModPhys.51.659,Wegner1971,FradkinShenker1979}. When 
coupled to staggered fermions, its entanglement structure becomes significantly richer, with 
string breaking leaving direct signatures in the entanglement entropy \cite{RevModPhys.51.659,Wegner1971,FradkinShenker1979,SenthilFisher2000,PhysRevB.84.235148,Zohar_2016,Kormos2017,PhysRevLett.118.266601,Smith2018,FRANK2020135484,PhysRevLett.124.120503,PhysRevLett.125.256401,Surace2020,Banuls2020,10.21468/SciPostPhys.10.6.148,Halimeh:2021lnv,Mildenberger:2022jqr,Kebric:2024yaa,Chen:2024oao,r6sr-dv13}. Defining entanglement entropy in LGTs is subtle because gauge constraints and a non-trivial centre prevent the physical Hilbert space from factoring into a simple tensor product, making the bipartition and hence the entropy inherently ambiguous~\cite{PhysRevD.89.085012}.
However tensor network methods particularly MPS provide an efficient numerical framework for 
studying its real-time dynamics~\cite{Pichler2016,Banuls:2018jag,Magnifico:2019kyj,Mathew:2025fim}. A parallel development concerns monitored quantum systems, where unitary evolution interspersed 
with projective measurements leads to measurement-induced phase transitions (MIPTs)~\cite{
skinner2019measurementinduced,li2018quantum,li2019measurementdriven,chan2019unitaryprojective}: 
a competition between entanglement-generating unitary dynamics (volume law) and 
information-extracting measurements (area law). Despite progress in both fields, the interplay 
between gauge constraints and quantum measurements remains largely unexplored.

In this work, we present preliminary results of a tensor network study of the $1+1$D $\mathbb{Z}_2$ 
gauge theory coupled to staggered fermions under local projective measurements 
in the no-click limit~\cite{PhysRevLett.68.580,PhysRevLett.70.2273,Plenio:1997ep} \footnote{A more detailed analysis including non-local filtering terms can be found in \cite{Chakrabarti:2026jjr}.}. We map 
out the entanglement dynamics as a function of measurement rate,  asking whether signatures of MIPTs emerge
and systematically compare the strong and weak coupling regimes, bridging high-energy 
phenomenology with the rapidly developing field of monitored quantum systems.

\section{Theoretical and computational frameworks}\label{section-1}
\subsection{Brief overview of 1+1D $\mathbb Z_2$ gauge theory}\label{section-1a}
%\subsection{ 1+1 dimensional $\mathbb Z_2$ gauge theory coupled with fermions}
We begin by briefly reviewing the necessary details regarding our model considered in this paper. We focus on a $\mathbb Z_2$ gauge theory coupled with dynamical staggered fermions in $1+1$ dimensions. The Hamiltonian for the system on the $L-$site lattice with open boundary condition is given as \cite{Wegner1971, FradkinShenker1979, Davoudi2024scatteringwave,Mildenberger2025}, 
\begin{equation}\label{H1}
 H = x\sum_{i=0}^{L-2} \left(\psi^{\dag}_{i}\tau^{X}_{i,i+1}\psi_{i+1}+ \text{h.c.} \right)
+ \mu \sum_{i=0}^{L-1} (-1)^{i}\psi^{\dag}_{i}\psi_{i}+ \sum_{i=0}^{L-2}\tau^Z_{i,i+1}\,, 
\end{equation}
where $\psi^{\dag}_{i} (\psi_{i})$ are fermionic creation (annihilation) operators acting on $i^{th}$ site 
%of the $\textrm{1d}$ lattice which have dynamical charges while
and $\tau^{\alpha}_{i,i+1}(\alpha=X,Z$ denotes the corresponding Pauli matrices) are the elements of $\mathbb Z_2$ representing the holonomies of gauge fields on each link between sites $i,i+1$.  In (\ref{H1}), the dimensionless parameters $x, \mu$ are related to the fermion mass $m$, coupling $g$ and lattice spacing $a$ as:
$$ x=\frac{1}{a^{2}g^{2}}, ~~\mu=2\frac{m}{g}\sqrt{x}\,.$$ For a fixed $\frac{m}{g}$ and lattice volume $V=La$, the continuum limit for the theory is obtained by taking $a\rightarrow 0$, $L\rightarrow\infty$ followed by $x\rightarrow\infty$. We will set $\frac{m}{g}=1$ for all of our computations. Hence, the coupling $x$ is the only tunable parameter in this case.

Being a gauge theory, the Hamiltonian is associated with the Gauss law operators  $G_i$ defined at each site $i$ as:
\begin{equation}
   G_i = \tau^Z_{i-1,i} \tau^Z_{i,i+1} e^{-i\pi \left( \psi_i^\dagger \psi_i - \frac{1 - (-1)^i}{2} \right)}\,.
\end{equation}
The Hamiltonian in (\ref{H1}) commutes with all the local Gauss law operators $[H,G_i]=0\,, \forall i$
and hence keeps the dynamics of the theory gauge-invariant.  The physical Hilbert space $\mathcal H_{phys}$ of the theory contains states $\{|\psi\rangle_{phys}\}$ which are gauge invariant, i.e $G_i\ket{\psi}=\ket{\psi}.$ A Jordan-Wigner transformation can be performed for the fermions, and the resulting Hamiltonian is given as: 
\begin{equation}\label{H2}
   H = x \sum_{i=0}^{L-2} \left( \sigma_i^{+} \, \tau^X_{i,i+1} \, \sigma_{i+1}^{-} + \text{h.c.} \right)
+ \mu\sum_{i=0}^{L-1}(-1)^{i}\sigma^{Z}_{i}+\sum_{i=0}^{L-2}\tau^   Z_{i,i+1}, 
\end{equation}
In (\ref{H2}), there exist two independent species of spins, $\tau$ corresponding to the two-component $\mathbb Z_2$ gauge fields defined on the links and $\sigma$ corresponding to fermions obtained via a Jordan-Wigner transformation. Both are represented by Pauli matrices. %\textcolor{red}{AB: define the $\sigma^{\pm}_i$ with appropriate factors of 2 in terms of Puali matrix.}
The physical Hilbert space is the gauge-invariant subspace of the full matter-gauge 
Hilbert space, with states satisfying the Gauss law constraint $G_i$ at each site. 
We adopt the convention that the presence (absence) of a fermion on an odd site, or 
absence (presence) of an anti-fermion on an even site, corresponds to spin up (down). 
Gauge invariance then requires the neighboring link spins to flip accordingly.

\begin{figure}[h]
    \centering
    \includegraphics[width=0.8\linewidth]{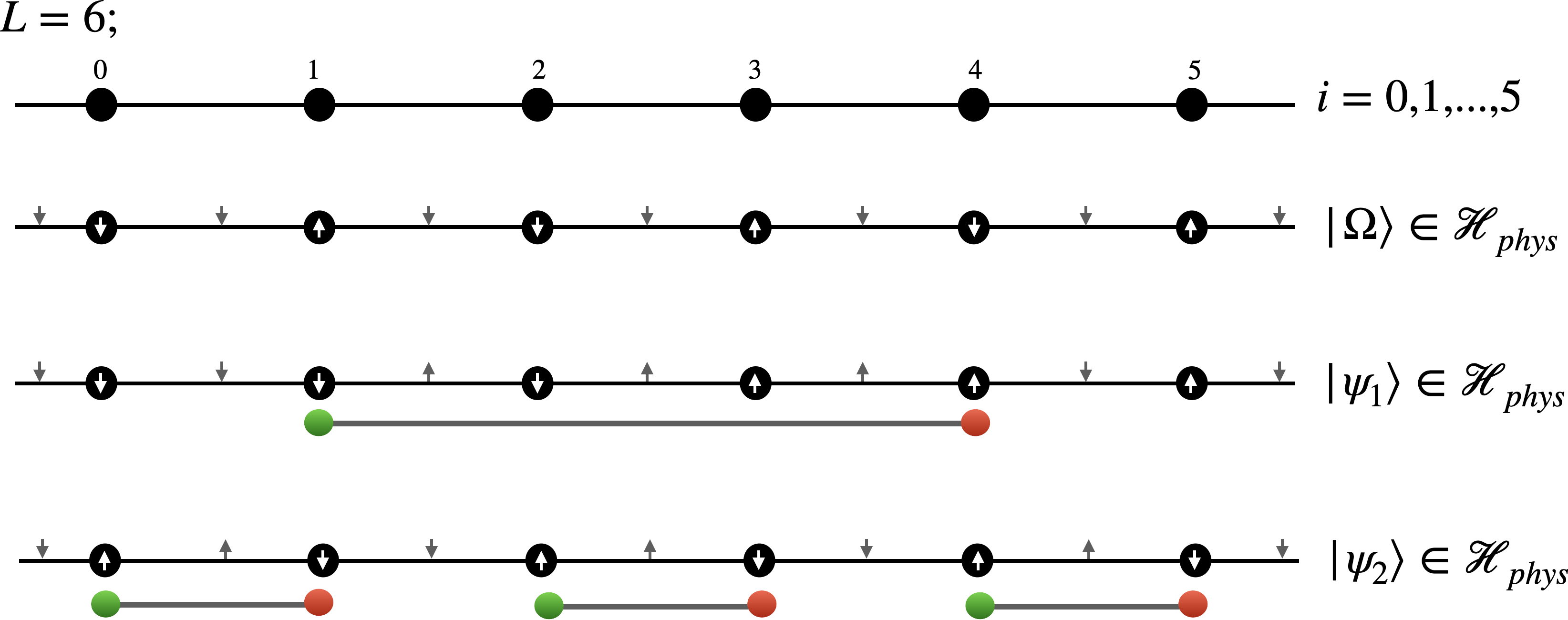}
    \caption{A diagram denoting physical states on a 6-site lattice. The topmost panel denotes the lattice and the label for each site;  the next one $|\Omega\rangle$ denotes the strong coupling vacuum - a global spin configuration on the lattice which corresponds to the presence of no particle, no anti-particle, and no gauge flux. The next two denote two global gauge invariant states $|\psi_1\rangle $ and $|\psi_2\rangle$, which contain 1 and 3 particle-antiparticle pairs, respectively, connected by gauge fluxes. Note that, for all the entanglement entropy configurations, the incoming boundary fluxes are chosen as zero ($s=-1/2$), and for both the outgoing ones are also zero ($s=-1/2$) as the difference between the total number of particles and anti-particles contained in the lattice is zero. All the entanglement entropy states correspond to a single global charge sector for the theory given by $s_{\textrm{global}}=-1/2$ \cite{Chakrabarti:2026jjr}.} 
    \label{fig-1}
\end{figure}

In this work, we consider quench dynamics from the \textit{strong coupling vacuum}: 
odd sites fully occupied by fermions, even sites empty, and no gauge flux on any link. 
In terms of Jordan-Wigner spins, this corresponds to all odd sites spin-up, all even 
sites spin-down, and all link spins at $s = -1/2$. Representative gauge-invariant 
states, including the strong coupling vacuum, are illustrated in 
Fig.~(\ref{fig-1}).

%The global matter-gauge Hilbert space is spanned by direct products of on-site spins and on-link spins over the entire lattice; however, the physical Hilbert space is a subset of that space in which all the states satisfy.
%As dictated by the operator $G_i$, the on-site contribution to a physical state is different for an even and an odd site where the fermions and anti-fermions reside. We work with the convention that the presence (absence) of a fermion (anti-fermion) is mapped to the spin to be up $s=1/2$ and absence (presence) of the same is mapped to the spin to be down $s=-1/2$. Gauge invariance at each site implies that the spins on the links are flipped in the presence (absence) of a fermion (anti-fermion). In this work, we consider quench dynamics starting from a gauge-invariant state. Most commonly, we consider the \textit{strong coupling} vacuum to be a gauge-invariant initial state. For a staggered theory, the strong coupling vacuum is defined as a configuration with odd sites fully filled with fermions and even sites all empty and no gauge flux present across the lattice. In terms of Jordan-Wigner spins, it is obtained as all odd sites to contain spin up and all even sites to contain spin down, while keeping all the on-link gauge fields at $s=-1/2$. A pictorial description of the strong coupling vacuum and a couple of other gauge invariant states is given in Fig.~(\ref{fig-1}).  

Now that we have discussed the underlying Hilbert space structure and the initial state on which we will apply our quench protocol, we will now discuss the exact protocol, which will include measurements. 

\subsection{A framework for monitoring Quantum Systems}
\begin{figure}[htb!]
\centering
\includegraphics[width=0.8\linewidth]{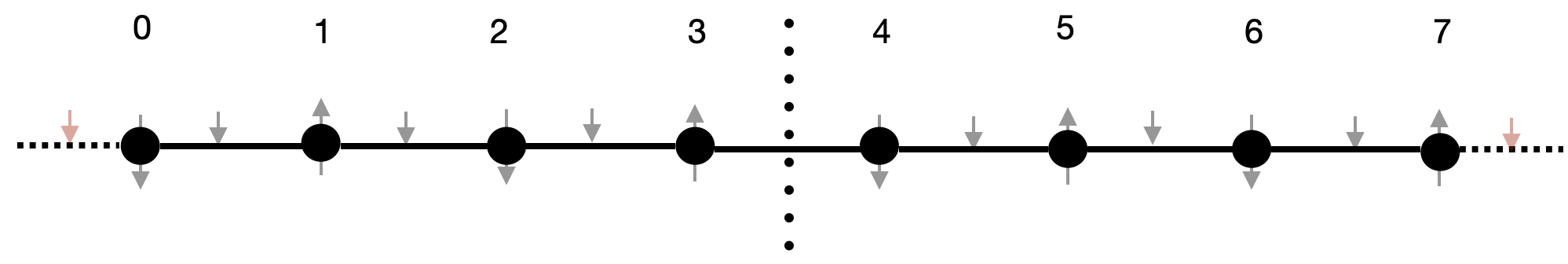}
\caption{\textcolor{black}{A representative diagram for a physical state on an 8-state lattice. The dotted line corresponds to the position of the link through which we bi-partition the system. The left side, consisting of 4 sites, is the subsystem $\textrm{A}$, and the right side is the subsystem $\textrm{B}$. When measuring local observables (either localized on each site or each link), we apply them to all sites of the system (in both the sub-systems A and B).}} 
%but for non-local observables (which acts on both links and sites), the results presented in the manuscript is mainly for the case where we restrict our measurements to one subsystem to avoid cutting through the link on whic the measurement operators are acting.}
%\textcolor{red}{NC: check the caption}
\label{fig for bipartition}
\end{figure}

We study entanglement dynamics under continuous monitoring in the \textit{no-click} 
limit~\cite{Wiseman2009,carmichael2009open,gardiner2004quantum,daley2014quantum,Jacobs2014,
PhysRevB.105.205125}, where the system evolves under an effective non-Hermitian Hamiltonian,
\begin{equation} \label{eq1}
    H_{\textrm{eff}} = H_0 - i\gamma H_1,
\end{equation}
with $H_0$ defined in~(\ref{H2}), $H_1$ encoding the measured observable, and $\gamma$ the 
measurement rate. This limit yields purely deterministic evolution, amenable to MPS methods, 
and has been shown to exhibit MIPTs in spin-chain models~\cite{10.21468/SciPostPhys.14.5.138,
Chakrabarti:2025hsb}. Starting from the strongly coupled vacuum $\rho(0)$, the state evolves as
\begin{equation}\label{denmat eqn}
    \rho(t) = \frac{e^{-iH\,t}\,\rho(0)\,e^{iH_{\textrm{eff}}^\dagger\,t}}
    {\textrm{Tr}\!\left(e^{-iH\,t}\,\rho(0)\,e^{iH_{\textrm{eff}}\,t}\right)},
\end{equation}
where the denominator ensures normalization at all times.
\textcolor{black} {We perform two types of local projective measurements where the projectors are constructed out of two physical observables: \textit{electric flux}, and \textit{particle-antiparticle density}}\footnote{\textcolor{black}{The $H_1$ term in (\ref{eq1}) for our case can be rewritten, up to an additive imaginary constant and a rescaling of $\gamma$, in terms of projectors. Thus, these cases involving local operators admit a standard no-click interpretation in terms of local loss channels.}}. The entanglement entropy of subsystem A, 
\begin{equation}
      S(L/2,t) = -\textrm{Tr}\,\Big(\rho_A(t)\ln\rho_A(t)\Big)
\end{equation}
  
 is computed from the reduced density matrix $\rho_A(t)$ obtained by tracing out subsystem B, 
as illustrated in Fig.~(\ref{fig for bipartition}). 
%Using $\rho_A(t)$, we compute the von Neumann entropy for subsystem A and denote it  $S(L/2,t)$. \begin{equation}S(L/2,t) =-\textrm{Tr}\Big(\rho_{A}(t)\ln \rho_A(t)\Big)\,. \label{ent}
%\end{equation} 

%As mentioned earlier, the Hamiltonian for the $1+1$D $\mathbb Z_2$ gauge theory coupled with a fermionic matter can be mapped to a spin model, which further simplifies the computation of entanglement entropy, as we can simply use (\ref{ent}). To investigate the presence of MIPT, we need to determine the scaling of $S(L/2,t)$ with system size, and it is crucial to probe systems of very large size. We achieve this by using tensor network tools as described below.

\subsection{Computational Framework}
We use Matrix Product States (MPS) via the ITensor package~\cite{itensor} for all numerical analysis. MPS represents quantum states as a chain of tensors, with entanglement controlled by the bond dimension. As long as entanglement remains limited, the computational cost scales polynomially with system size.
%Ground states are obtained via DMRG using the Matrix Product Operator (MPO) form of the Hamiltonian~\cite{Orus:2013kga, Bridgeman:2016dhh},
Time evolution is performed using the \textcolor{black}{second-order TEBD method}~\cite{itensor} and entanglement entropy is extracted efficiently via singular value decomposition (SVD) at the central bond, giving the Schmidt coefficients  directly, from which the von Neumann entropy is computed. 
\begin{figure}
    \centering
    \begin{subfigure}[b]{0.48\textwidth}
        \centering
        \includegraphics[width=\linewidth]{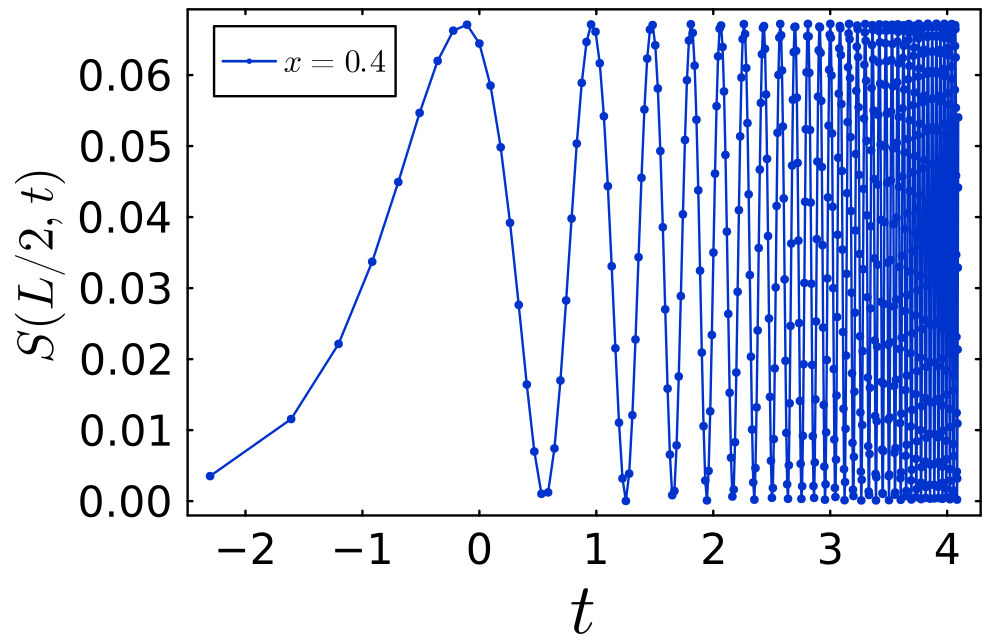}
        \caption{}
      
    \end{subfigure}
    \begin{subfigure}[b]{0.48\textwidth}
        \centering
        \includegraphics[width=\linewidth]{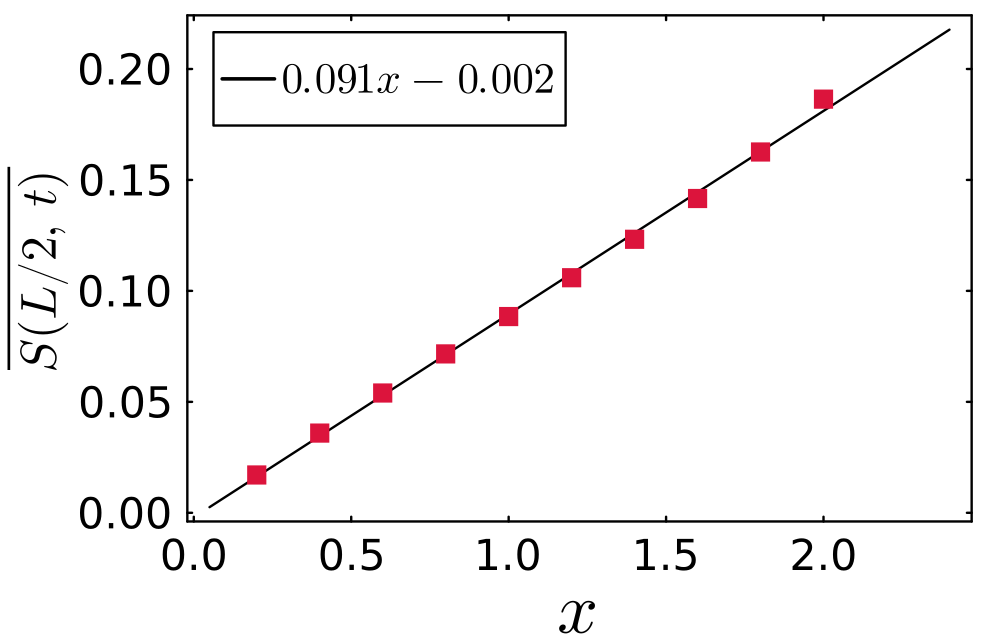}
        \caption{}
       
    \end{subfigure}
    \caption{(a) The time evolution of entanglement entropy without any measurement clearly shows that entanglement entropy does not show any saturation even at late time, (b) time-averaged entanglement entropy \cite{Chakrabarti:2026jjr}. }
    \label{fig1}
\end{figure}

%\subsection*{Dynamics of Entanglement entropy of $\mathbb Z_2$ gauge theory without any measurement}\label{section-3}

%Time evolution is performed using a second-order Trotter decomposition on 
%non-overlapping three-site clusters, with local truncation error $\mathcal{O}(\delta^3)$ 
%per step, as implemented in ITensors~\cite{itensor}.
We benchmark our tensor network code against an exact-diagonalization code for a small system size.  We explored bond dimensions up to $\chi = 1000$ to ensure a $10^{-8}$
convergence of entanglement entropy.
We work with a system of $L$ sites with 
time step $\delta = 0.1$, total evolution time $T = 60$, starting from the strongly coupled vacuum with 
open boundary conditions. Gauss's law is explicitly verified throughout the evolution to 
preserve the gauge structure. As shown in Fig.~(\ref{fig1}) (a), the entanglement entropy grows and oscillates at late times across different values of $x$ without saturating, suggesting the absence of thermalization in the 
$1+1$D $\mathbb{Z}_2$ theory without any measurements. The time-averaged entanglement entropy increases monotonically with $x$, consistent with growing entanglement towards the continuum limit as shown in Fig.~(\ref{fig1}) (b) .

\section{Results: Measuring local physical observables in $\mathbb{Z}_2$ gauge theory}\label{section-4}
\begin{figure}
    \centering
    \begin{subfigure}[b]{0.50\textwidth}
        \centering

        \includegraphics[width=\linewidth]{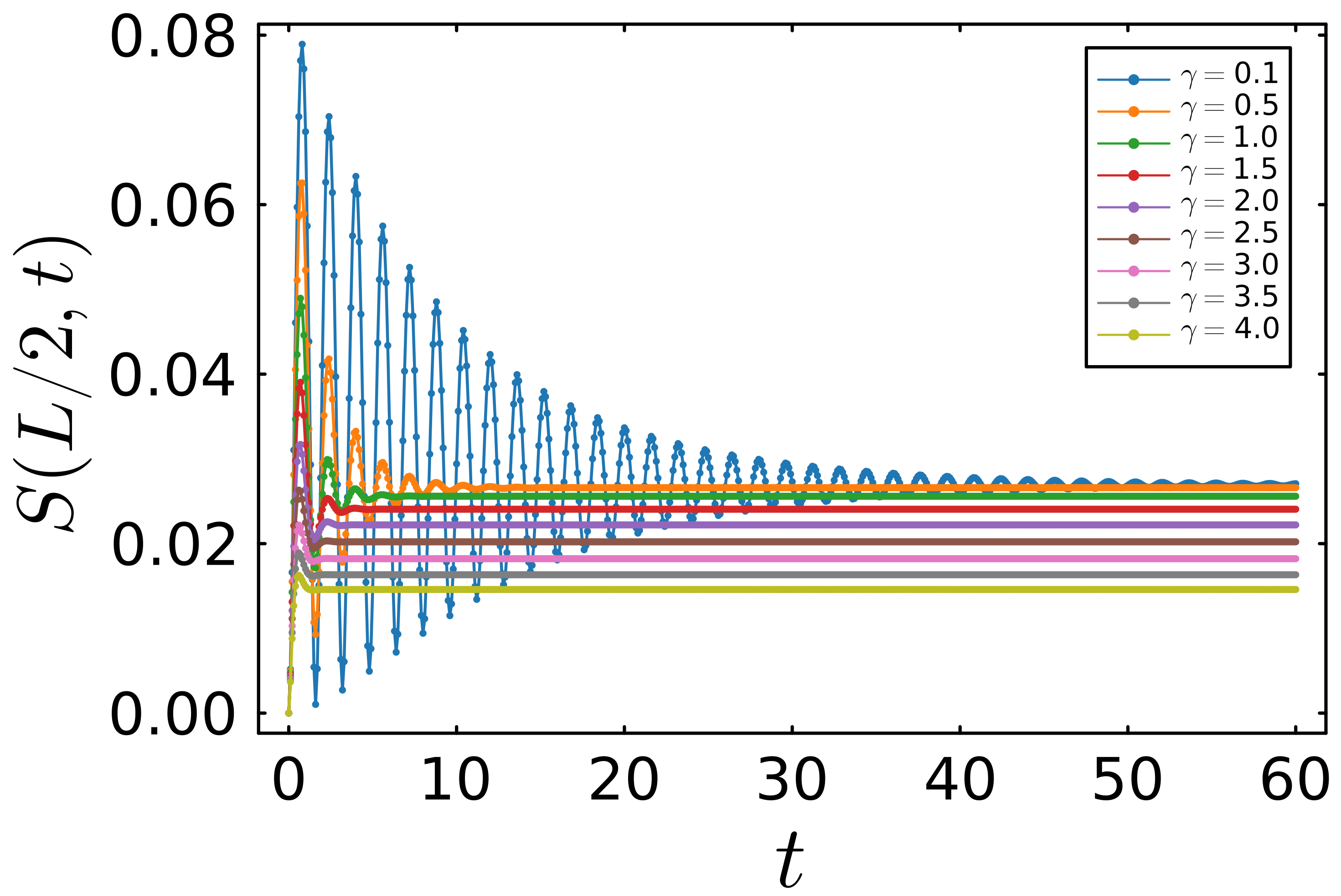}
        \caption{}
        
    \end{subfigure}
    \begin{subfigure}[b]{0.42\textwidth}
        \centering
        \includegraphics[width=\linewidth]{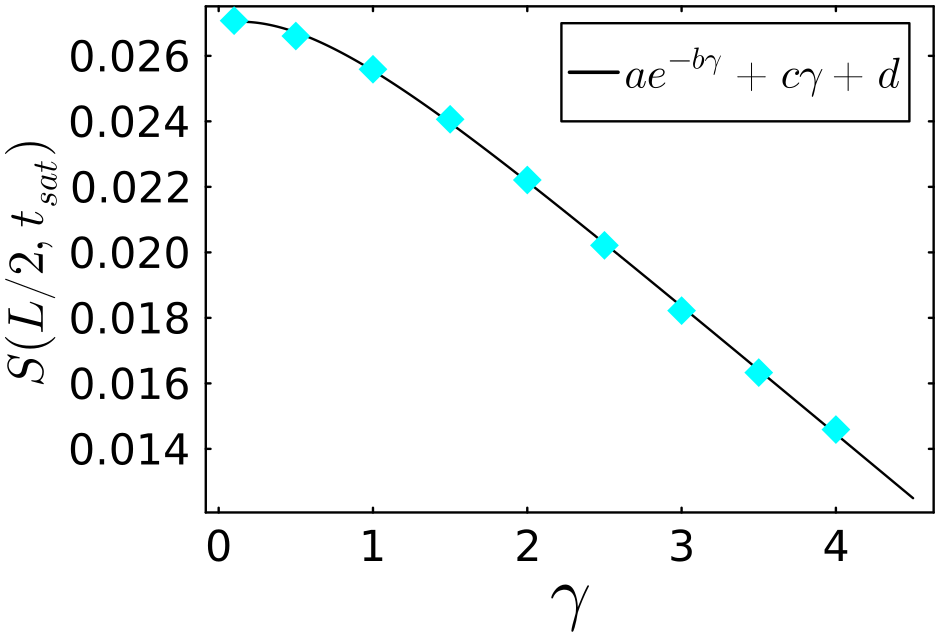}
        \caption{}
        
    \end{subfigure}
    \caption{(a) Entanglement dynamics under the measurement of the electric flux operator for
different measurement rates for $x = 0.5$ and $L = 64\,.$ In (b) \textcolor{black}{we show the saturation value of EE as a function $\gamma$ $(a=0.9735,~b = 0.0358,~c =0.0267,~d = -0.9447)$ evaluated at  $t_{sat}= 60\,.$} }
    \label{fig2}
\end{figure}

\begin{figure}
    \centering
    \begin{subfigure}[b]{0.50\textwidth}
        \centering
        \includegraphics[width=\linewidth]{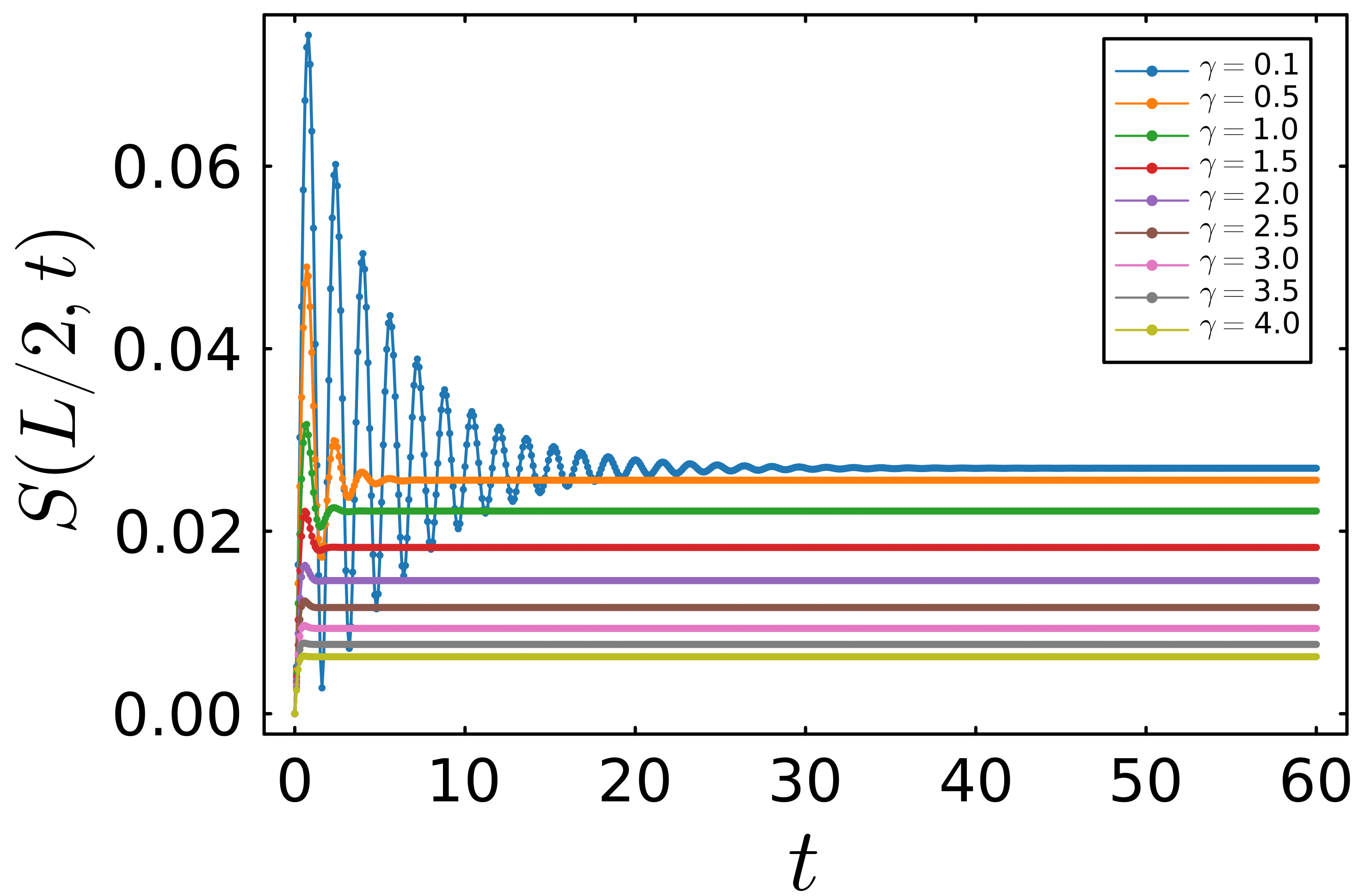}
        \caption{}
        
    \end{subfigure}
    \begin{subfigure}[b]{0.42\textwidth}
        \centering
        \includegraphics[width=\linewidth]{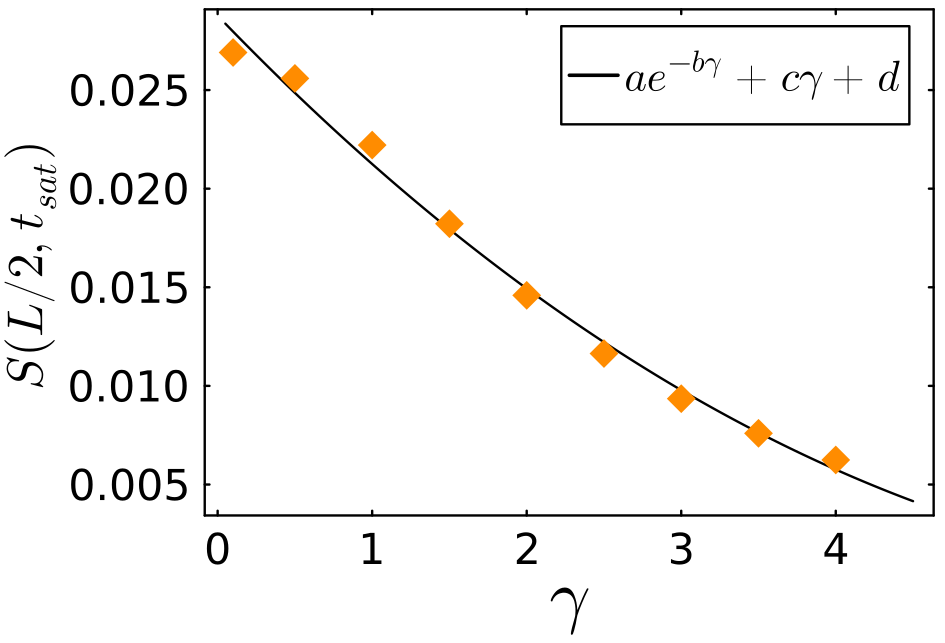}
        \caption{}
        
    \end{subfigure}
    \caption{(a) Entanglement dynamics under the measurement of the particle-anti operator for
different measurement rates for $x = 0.5$ and $L = 64\,.$ In (b)
\textcolor{black}{We show saturation value of EE as a function of $\gamma$  $(a=-0.0032,~b = 1.6452,~c =-0.0039,~d = 0.0301)$ evaluated at  $t_{sat}= 60$}. }
    \label{fig3}
\end{figure}

\begin{figure}
    \centering
    \begin{subfigure}[b]{0.48\textwidth}
        \centering
        \includegraphics[width=\linewidth]{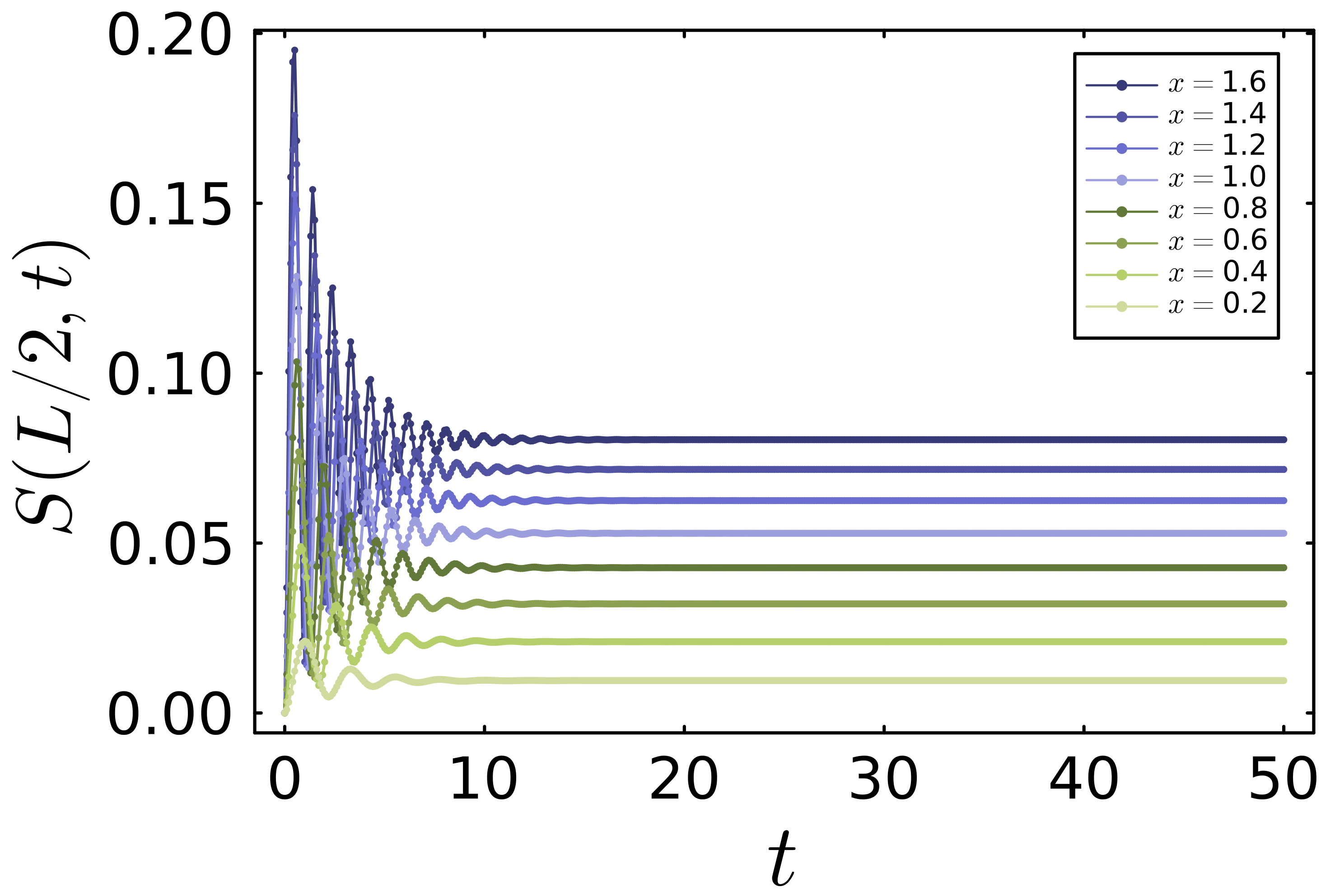}
        \caption{}
       
    \end{subfigure}
    \begin{subfigure}[b]{0.48\textwidth}
        \centering
        \includegraphics[width=\linewidth]{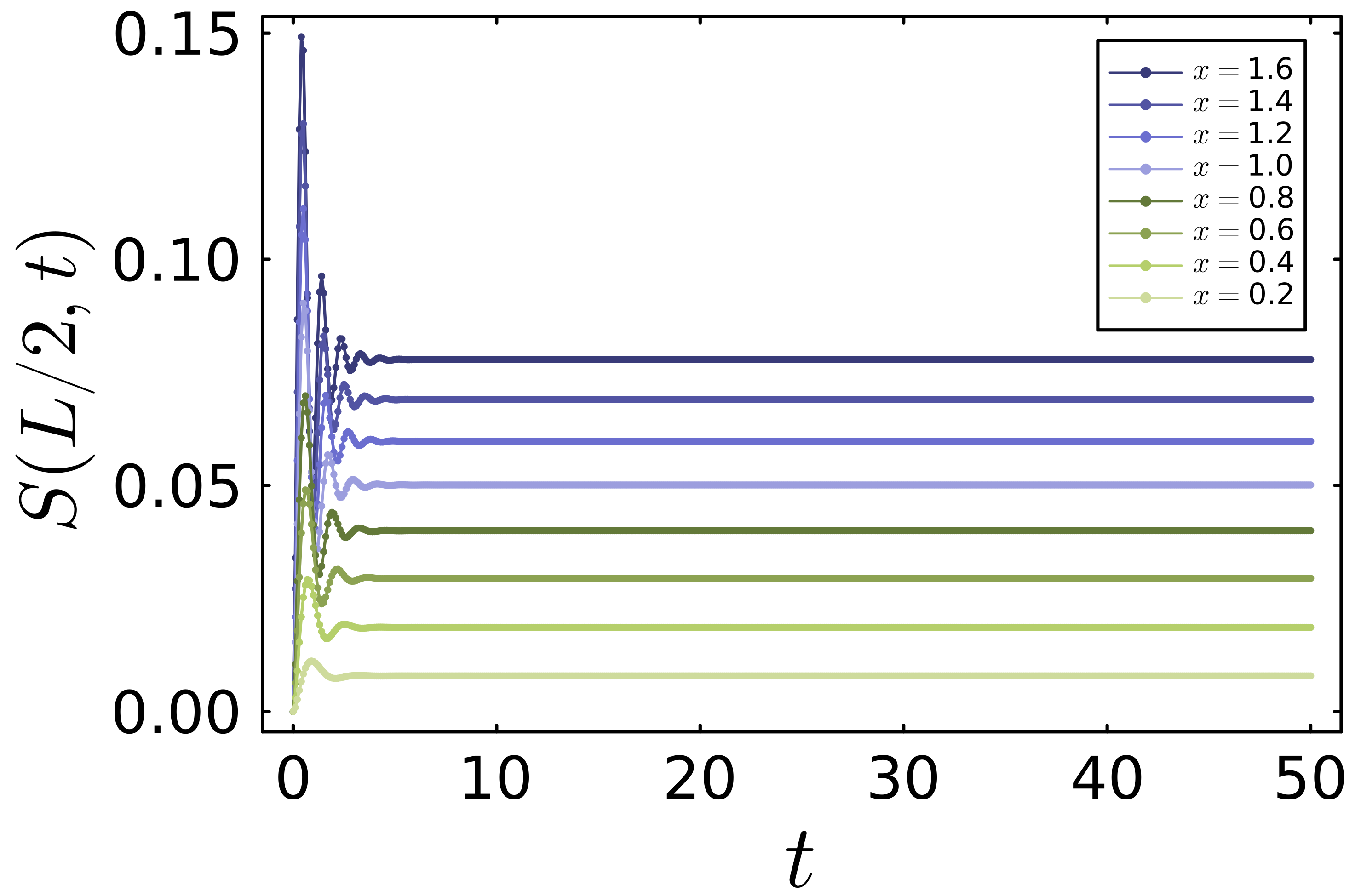}
        \caption{}
      
    \end{subfigure}
    \caption{(a) Entanglement dynamics under the measurement of the electric flux operator for
different $x = 0.5$ and $L = 64$ with (a) $\gamma=0.5$ and (b) $\gamma=1.5\,.$}
    \label{fig4}
\end{figure}

The objective of this study is to find the dynamics of entanglement in the presence of measurement for $\mathbb{Z}_{2}$ gauge Hamiltonian. Note that, in this work, we have chosen the Hamiltonian (\ref{H1}) to be represented in the electric field basis. Hence, the simplest measurement in this framework is the measurement of electric flux (contributing to electric energy) and the fermionic staggered mass (contributing to the mass energy) of the system. 
%Here, we use a non-Hermitian Hamiltonian, which is derived as the no-click limit of the Stochastic Schrodinger equation, where the noise components are excluded, resulting in a purely deterministic evolution. 
While measuring those, we turn our attention to the effect of measurement on the dynamics of entanglement entropy. The main motivation is twofold: (a) whether the entanglement saturates under the measurement, and (b) whether we observe a \textcolor{black}{MIPT-like transition}. We measure two different physical observables: electric flux and particle-antiparticle number.
%and mesonic excitations. While the first two are local operators, the last one is a non-local operator. Our setup provides a good opportunity to discuss the dynamics of entanglement entropyunder non-local measurements, which have well-defined physical meaning, while most previous studies in the context of MIPT are largely confined to local measurements on spin chains or circuits. Hence, this makes the current study all the more exciting. In the next section, we investigate the effect of measuring local operators. 
We plot the evolution of entanglement entropy over time for various measurement rates for measuring electric flux in Fig~\ref{fig2}(a) and staggered fermion number or fermionic mass in Fig~\ref{fig3}(a). Several features emerge from these plots; we list them below:
\begin{enumerate}

\item Unlike the case with no measurement, the entanglement entropy saturates at late times after 
early-time oscillations, the saturation value decreasing monotonically with 
the measurement strength $\gamma$. A comparison of the saturation values across the measurement 
rates reveals qualitative differences between the two local measurement operators (for a fixed $x$) as illustrated in Fig~\ref{fig2}(b) and Fig~\ref{fig3}(b).

\item We fix the measurement rate and compute the entanglement entropy for different values of the coupling parameter $x$ while measuring the local observables. From Figs~(\ref{fig4}) (a) and (b), we can observe that with increasing $x$, the late time saturation value of the entanglement entropy is increasing linearly with the coupling $x\,.$ This behavior remains the same if we measure local observables for different values of the measurement parameter $\gamma$.

\item We further find that the dynamics of entanglement entropy is independent of the
sub-system size. We have exhaustively scanned over values of $\gamma$ for fixed values of $x$, and vice versa.
This remains robust. \textcolor{black}{Hence, there is no MIPT-like transition (at least within the no-click limit) for this system under the measurement of the local operators across the range of measurement strengths, evolution times, and system sizes considered here}.
\end{enumerate}
%\begin{equation}
%\label{H3}
%   H_{\textrm{eff}} = x \sum_{j=0}^{L-2} \left( \sigma_j^{+} \, \tau^X_{j,j+1} \, \sigma_{j+1}^{-} + \text{h.c.} \right)
%+ \mu\sum_{j=0}^{L-1}(-1)^{j}\sigma^{Z}_{j}+\sum_{j=0}^{L-2}1\tau^Z_{j,j+1}-i\gamma\mathcal{H}
%\end{equation}

\section{Conclusion}
We have studied entanglement dynamics in $1+1$D $\mathbb{Z}_2$ gauge theory coupled to 
staggered fermions under continuous monitoring in the no-click limit, with Gauss's law 
preserved throughout. To our knowledge, this is the first study of measurement-induced 
entanglement dynamics in a lattice gauge theory. Without measurement, entanglement entropy grows and oscillates without saturating, with the time-averaged 
entanglement entropy increasing with the coupling $x$. Under local measurements, it saturates at late times in 
all cases studied. For both local observables (electric flux and particle-antiparticle 
density) the late-time saturation value is 
independent of system size for all $\gamma$, indicating the \textit{absence of MIPT}-like transition 
in the no-click limit. For \textcolor{black}{local projective measurements}, entanglement entropy saturates after early oscillations 
that diminish with increasing $\gamma$, consistent with quantum Zeno behavior.
Future directions include studying non-local measurements, extending beyond the no-click limit to incorporate stochasticity, developing quantum circuit realisations for hardware implementation, and generalising to continuous and non-Abelian gauge theories such as $SU(2)$ loop-string-hadron Hamiltonian description \cite{Raychowdhury:2019iki}, using the tensor network calculation \cite{Mathew:2025fim, Gupta:2026tcg} as well as its quantum circuit implementation as in \cite{Ilcic:2026cac}.

% ------------------------------------------------------------------

% ------------------------------------------------------------------
\acknowledgments
Research of IR is supported by the  OPERA award (FR/SCM/11-Dec-2020/PHY) from BITS-Pilani, the Start-up Research Grant (SRG/2022/000972) from ANRF, India and the cross-discipline research fund (C1/23/185) from BITS-Pilani. IR acknowledges discussions at QC4HEP working group meetings. NN is supported by the CSIR fellowship
provided by the Govt. of India under the CSIR-JRF scheme (file no. 09/1031(19779)/2024-
EMR-I). NC is supported by the Director’s Fellowship of the Indian Institute of Technology
Gandhinagar. AB would like to thank the speakers and participants of the BIRS-CMI workshop (25w5386) ``Quantum Gravity and Information Theory: Modern Developments" for stimulating discussion, and the physics department of BITS-Pilani, Goa, for hospitality during this work. AB is supported by the Core Research Grant (CRG/2023/ 001120) by the Anusandhan National Research Foundation (ANRF), India.  A.B. also acknowledges the associateship program of the Indian Academy of Sciences, Bengaluru, and support from the Indian Institute of Technology Gandhinagar and a generous donor through the Singheswari and Ram Krishna Jha Chair.  NA would like to thank the Start-up Research Grant (SRG/2022/000972) for the computational facilities at BITS, and Sharanga HPC usage for tensor network computations. NA further acknowledges Sharanga HPC usage for tensor network computations, Kapil Ghadiali for computational support at TIFR and hospitality at the Institute for Basic Science, Daejeon, where a part of the work was done. 

% ------------------------------------------------------------------

% ------------------------------------------------------------------
\bibliographystyle{JHEP}
\bibliography{proc}
\end{document}